\documentclass[letterpaper]{article}
\usepackage[T1]{fontenc}
\usepackage{spconf,amsmath,amssymb,graphicx,booktabs,array}
\usepackage[hidelinks]{hyperref}
\usepackage{placeins,balance}
\title{Beyond Sparsity: Weight Location and Network Context in Pruned MRI Reconstruction}
\name{Mohammed Wattad, Tamir Shor, and Alexander M. Bronstein}
\address{Faculty of Computer Science, Technion - Israel Institute of Technology, Haifa, Israel}
\begin{document}
\ninept
\maketitle
\begin{abstract}
Pruning reduces the number of weights in magnetic resonance imaging (MRI) reconstruction networks. Equal sparsity, however, can retain weights with different computational roles and different compatibility with the trained network. We study these effects across 120 convolutional U-Net and vision transformer models using controlled edits evaluated before retraining. In U-Net, preserving high-resolution computation improves reconstruction at equal deletion counts across all tested sparsity levels. Equal-operation controls reveal additional sensitivity of the first convolution, which operation count alone cannot explain. In transformers, retained pretrained weight energy correlates with quality within sparsity levels, yet the same masks reverse their reconstruction ranking when the surrounding trained weights change. At 90\% sparsity, matching intermediate output scales largely removes a high-energy-mask penalty while retaining an advantage for the network's original mask. Thus sparsity alone does not characterize reconstruction quality: architecture-specific descriptors are informative, but mask quality can still depend on the surrounding trained network.
\end{abstract}

\section{Introduction}
Magnetic resonance imaging (MRI) reconstruction estimates an image from spatial-frequency measurements, called k-space. Acquiring fewer measurements can shorten a scan, while increasing the reconstruction task's reliance on learned image structure. Pruning can reduce the storage and theoretical computation of reconstruction networks, but its effect on image quality depends on which connections remain. Understanding that dependence is particularly relevant when acquisition conditions differ from training~\cite{wattad_kspace,antun,robustness,mri_generalization}.

A pruning mask is a binary array that keeps selected weights and sets the others to zero. Sparsity is the fraction of eligible weights removed. Equal sparsity provides a natural comparison of model size, yet weights can play different roles. A convolutional weight is reused at every spatial position of its feature map. A weight in a high-resolution layer therefore contributes to more computations than one in a low-resolution layer. In a vision transformer (ViT), the pruned multilayer perceptron (MLP) weights all process the same image tokens, so their reuse is uniform. In both architectures, their effect also depends on the trained parameters that receive and process their outputs.

Prior pruning methods use second-order or loss/gradient criteria~\cite{obd,snip,grasp}, data-free synaptic-flow criteria~\cite{synflow}, layer-adaptive magnitude scores~\cite{lamp}, activation-aware scores~\cite{wanda}, or dynamic connectivity during training~\cite{rigl}. PUN-IT learns masks over fixed weights and has been studied for MRI generalization~\cite{punit}. Sparse-network performance also depends on architecture and initialization~\cite{lottery,rethinking,missing}. Prior pruning work has emphasized that parameter count and FLOPs measure different aspects of efficiency~\cite{blalock}. Here we investigate which structural effects remain after matching counts, and whether a mask's reconstruction advantage persists in another trained network.

We combine observations across 120 trained models with controlled edits to test three aspects of pruning. (i) Within sparsity levels, we identify architecture-specific associations with reconstruction quality: operation count in U-Net and retained pretrained energy in ViT. (ii) Equal-count U-Net deletions demonstrate sensitivity to spatial allocation, while equal-operation controls isolate an additional first-convolution effect. (iii) ViT mask exchanges establish context-dependent rankings, and output-scale calibration implicates scale mismatch in a high-energy-mask penalty at extreme sparsity. The main edits are evaluated before retraining to expose their immediate effect on a fixed network.

\section{Analysis}
To distinguish storage from arithmetic, let $a_l$ denote the number of active weights in layer $l$, $u_l$ their number of uses, and $F_{\rm fixed}$ the operations outside the pruned layers. Counting a multiply-add as two operations gives the theoretical cost in billions of floating-point operations (GFLOPs):
\begin{equation}
 F_{\rm eff}=10^{-9}\left(F_{\rm fixed}+2\sum_l u_l a_l\right).
 \label{eq:reuse}
\end{equation}
In our U-Net, feature maps range from $320\times320$ to $20\times20$, giving a 256-fold difference in weight reuse. Every pruned ViT MLP matrix processes 1,024 tokens. Thus equal weight counts can hide different U-Net operation counts, whereas count-matched ViT masks have equal theoretical cost. Equation~\ref{eq:reuse} assumes that zero weights can be skipped. It excludes Fourier transforms and memory access, and does not measure inference time.

To compare which values a mask preserves, let $W_l^0$ be the shared pretrained ViT MLP matrix and $M_l$ its binary mask. With $\odot$ denoting elementwise multiplication, define retained energy as
\begin{equation}
 E(M)=\frac{\sum_l\|M_l\odot W_l^0\|_F^2}
 {\sum_l\|W_l^0\|_F^2}.
 \label{eq:energy}
\end{equation}
The squared Frobenius norm sums squared entries. We evaluate $f(B,M\odot W^0)$, where the network context $B$ comprises trained attention, embeddings, biases, normalization and output layers. Each mask retains the same locations, values and energy as $B$ changes. A reversal in the quality ordering of two masks therefore establishes a dependence on context that a fixed mask-only ranking cannot capture.

\par
For a complementary energy test, we construct masks with the same active count as the context's original mask $M_{\rm own}$ and zero overlap with it:
\begin{equation}
 \mathcal M=\{M:\|M\|_0=\|M_{\rm own}\|_0,\ M\odot M_{\rm own}=0\}.
 \label{eq:built}
\end{equation}
Here $\|M\|_0$ counts active entries. Selecting available pretrained weights by low magnitude, uniformly at random, or by high magnitude varies energy at fixed count and overlap. Weight locations and layer allocation can still change, so this construction tests energy-directed selection under those controls.

\section{Setup}
We train on fastMRI brain~\cite{fastmri}, using 4,671 training slices from 527 files and 2,814 validation slices from 320 files with disjoint patient IDs, and evaluate on separate fastMRI test volumes and low-field M4Raw validation data~\cite{m4raw}. Each architecture has 12 models at each of 50, 60, 70, 80 and 90\% sparsity: five image-SNIP masks~\cite{snip}, five random masks, one PUN-IT and one RigL. Image-SNIP uses loss sensitivity on one training slice per mask. After mask selection, retained weights and unpruned parameters are trained. RigL updates connectivity during training. Sparsity refers to eligible weights: convolutional and linear weights in U-Net~\cite{unet}, and 20 MLP matrices in ViT~\cite{vit}. At 70\% MLP sparsity, 54.06\% of all ViT parameters remain active, because attention and other components remain dense.

The U-Net has four pooling stages and 3.35 million parameters. ViT has ten blocks, 704-dimensional tokens, $10\times10$ patches and 60.4 million parameters. It uses a shared reconstruction-pretrained checkpoint as the reference $W^0$ for all mask comparisons. U-Net selection and training initialize separately. Adam minimizes mean absolute pixel error with batch size one and training seed 42. U-Net trains for 50 epochs at peak learning rate $5\times10^{-4}$, ViT for 40 at $10^{-4}$. Validation peak signal-to-noise ratio (PSNR, higher is better) selects checkpoints.

For each coil image, we combine real components and imaginary components separately by root-sum-of-squares, then form a complex image from those two results. Sampling its Fourier transform is simulated, retaining about one quarter of frequency lines, including a central fraction of .08. This construction differs from reconstructing native acquired multicoil data. The input is the magnitude of the inverse Fourier transform after missing measurements are zero-filled. Input and target are independently normalized to $[0,1]$. Sampling jitter shifts lines by rounded Gaussian offsets with standard deviation three lines, rejecting collisions and out-of-range moves. We evaluate 1,500 slices per dataset with and without jitter. Scores average slice PSNR in decibels (dB), using each target's intensity range.

The initial controlled study uses four models at 70\% sparsity, one per method. Expanded tests use all 12 models at the specified levels. Reported adjusted intervals use 40,000 paired bootstrap resamples of 170 fastMRI volumes or 23 M4Raw study-ID groups, with Bonferroni adjustment within each original family of 16 comparisons. U-Net scores first average five deletion seeds per image. Intervals describe evaluation-group variation for fixed networks. Expanded sweeps and correlations are descriptive.

\begin{table}[t]
\centering
\caption{Within-sparsity Pearson correlations with PSNR. Entries span five levels, with 12 trained models per level. ViT energy uses Eq.~\ref{eq:energy}. These are descriptive associations.}
\label{tab:descriptors}
\setlength{\tabcolsep}{4pt}
\begin{tabular}{@{}llcc@{}}\toprule
Model & Descriptor & fastMRI & M4Raw\\\midrule
U-Net & Active count & $-.17$ to $.39$ & $-.34$ to $.19$\\
 & Operations & $.54$ to $.72$ & $.14$ to $.69$\\
ViT & Energy & $.81$ to $.97$ & $.77$ to $.96$\\
\bottomrule\end{tabular}
\end{table}

\begin{table}[t]
\centering
\caption{U-Net PSNR difference H minus L (dB). H deletes at low resolution first, L at high resolution first, with equal counts. Each range spans 12 models after averaging five deletion seeds. F denotes fastMRI, M denotes M4Raw, and J denotes sampling jitter. All 240 means are positive.}
\label{tab:deletion}
\setlength{\tabcolsep}{3pt}
\begin{tabular}{@{}ccccc@{}}\toprule
Sparsity & F & F+J & M & M+J\\\midrule
50\% & 3.49-8.72 & 1.28-4.34 & 1.28-4.22 & .81-3.01\\
60\% & 2.30-9.20 & .86-5.17 & .64-4.44 & .35-3.45\\
70\% & 1.52-8.07 & .36-4.41 & .85-4.21 & .43-3.01\\
80\% & 2.11-7.66 & .06-4.22 & .95-3.74 & .53-2.89\\
90\% & 2.35-7.89 & .79-4.48 & .87-4.33 & .48-3.27\\
\bottomrule\end{tabular}
\end{table}

\section{Results}
\subsection{U-Net Results}
\begin{figure}[!b]
\centering
\includegraphics[width=\columnwidth]{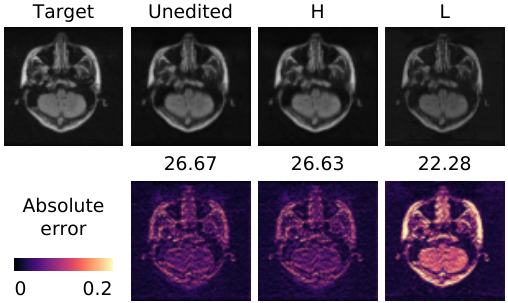}
\caption{M4Raw reconstruction from a SNIP U-Net at 70\% sparsity, using the first deletion seed. H preserves high-resolution computation, while L deletes it first. The slice was selected before visual inspection as the example closest to the median H-minus-L PSNR difference. Labels give PSNR in dB. Image intensities and absolute errors relative to the target use shared ranges $[0,1]$ and $[0,.2]$.}
\label{fig:recon}
\end{figure}

\begin{figure*}[t]
\centering
\includegraphics[width=\textwidth]{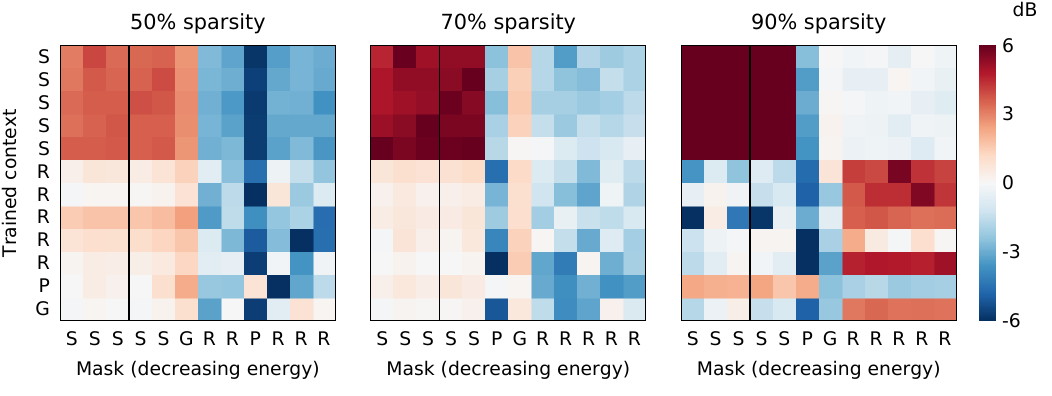}
\caption{ViT mask exchange on M4Raw, with 144 combinations per sparsity level. Rows select trained context and columns select masks in decreasing retained energy. S denotes SNIP, R random, P PUN-IT, and G RigL. Color shows PSNR relative to each context's median across 12 masks, clipped at $\pm6$ dB. The line separates the three highest-energy masks. High-energy masks benefit SNIP contexts, while their preference reverses in random contexts at 90\% sparsity. All MLP values are reset to the shared pretrained reference.}
\label{fig:exchange}
\end{figure*}

We first ask whether masks at similar sparsity differ in ways that track quality. Within each sparsity level, operation count has a positive association with U-Net PSNR, while retained weight count has weak associations of varying sign (Table~\ref{tab:descriptors}). Count is nearly fixed by design, so its within-level correlations reflect only small achieved-count differences. The operation-count association motivates an intervention on spatial allocation. Associations alone cannot distinguish weight reuse from the roles of individual layers. The same qualitative separation holds for structural similarity (SSIM)~\cite{ssim}: operation count remains positively associated with quality, while within-level active-count associations are weak.

For equal-count deletions, rule H preserves high-resolution computation by deleting from low-resolution layers first. Rule L deletes from high-resolution layers first. Each removes 10\% of active weights in the 18 body convolutions, capped at 20\% of the active filter weights per output channel. The three output convolutions remain unchanged. Five paired random seeds order deletions within each resolution, and surviving values remain fixed.

H has higher PSNR in all 240 model-condition comparisons across five sparsity levels (Table~\ref{tab:deletion}). The initial 16 method-condition comparisons at 70\% also have adjusted intervals above zero. The deleted weights have similar magnitudes under H and L, while their reuse-weighted squared magnitudes differ by more than an order of magnitude. These results establish an effect of deletion allocation, which changes both reuse and layer function.

To separate these factors, we compare deletions at the same spatial resolution, matching both count and operation cost. Table~\ref{tab:location} reports these controls on M4Raw at 70\% sparsity. Moving deletions from the first convolution to its adjacent convolution improves scale-invariant PSNR by 1.09 to 2.87 dB in criterion-based models. Equal-resolution encoder-decoder differences are smaller. Four random-mask models have a first-layer difference of only .05 to .13 dB. The fifth retains a dense first layer and is excluded from that comparison. Grouped-bootstrap intervals for the matched first-convolution comparisons exclude zero for the criterion-based models. The effect therefore depends on the learned sparse network, as well as the layer's position.

The scale-invariant scores remove a global output-scale difference before evaluating image error. The raw first-layer penalties are 19 to 46\% larger, so output scale explains part of the sensitivity. The remaining effect at equal count and operations shows why a computational descriptor alone is insufficient.
\begin{table}[t]
\centering
\caption{Equal-count, equal-operation U-Net deletions on M4Raw at 70\% sparsity. Entries are scale-invariant PSNR differences (dB). First minus adjacent compares the first convolution with its neighbor. Encoder minus decoder spans four matched resolutions. Negative means the first-named location is more damaging. Ranges span masks and, in the last column, resolutions.}
\label{tab:location}
\setlength{\tabcolsep}{5pt}
\begin{tabular}{@{}lcc@{}}\toprule
Model family & First minus & Encoder minus\\
 & adjacent & decoder\\\midrule
PUN-IT & $-1.43$ & $-.05$ to $.08$\\
RigL & $-1.09$ & $-.04$ to $.01$\\
SNIP & $-2.87$ to $-1.35$ & $-.18$ to $.14$\\
Random & $-.13$ to $-.05$ & $-.34$ to $.09$\\
\bottomrule\end{tabular}
\end{table}

An exploratory rule $L_1$, designed after inspecting outputs, protects the first convolution and relocates its planned deletions. At less than .05\% change in operation count, it recovers a median 87 to 93\% of the H-minus-L gap in non-random models at each sparsity level. Figure~\ref{fig:recon} illustrates the H-versus-L difference on a slice selected before visual inspection. Together, these controls show that operation count is informative, while the first convolution contributes additional sensitivity. Its role in processing the aliased input is a possible explanation that these experiments do not isolate.

\begin{table}[t]
\centering
\caption{SNIP-mask minus RigL-mask PSNR on M4Raw at 70\% ViT sparsity. All masks use shared pretrained MLP values and 94.935 GFLOPs. Brackets give group-bootstrap intervals adjusted within 16 comparisons. Positive favors SNIP.}
\label{tab:context}
\setlength{\tabcolsep}{7pt}
\begin{tabular}{@{}lrr@{}}\toprule
Trained context & Difference (dB) & Interval\\\midrule
SNIP & $+4.243$ & $[4.034,4.458]$\\
Random & $-.177$ & $[-.566,.235]$\\
PUN-IT & $+.513$ & $[-.034,1.098]$\\
RigL & $-.474$ & $[-.847,-.061]$\\
\bottomrule\end{tabular}
\end{table}

\subsection{ViT Results}

\begin{figure*}[t]
\centering
\includegraphics[width=.91\textwidth]{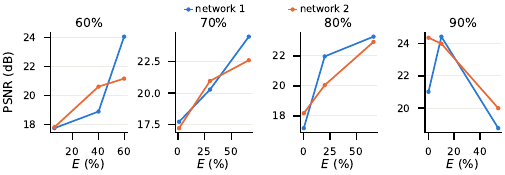}
\caption{Controlled retained-energy test on M4Raw. Masks keep equal counts and have zero overlap with the context's original mask (Eq.~\ref{eq:built}). Each line evaluates low-energy, random and high-energy selections in one of two randomly pruned contexts. Fitted energy-quality slopes are positive at 60 to 80\% sparsity and negative at 90\%. Each line contains only three masks, and the middle-energy mask is best in one 90\% context.}
\label{fig:energy}
\end{figure*}
Equal reuse makes operation count constant for count-matched ViT masks. Retained energy nevertheless correlates with PSNR within every sparsity level on both datasets (Table~\ref{tab:descriptors}). To test whether this association implies a transferable mask ranking, we exchange all 12 masks among all 12 contexts at 50, 70 and 90\% sparsity. Every combination, including the original-mask baseline, resets MLP values to $W^0$, so masks bring only locations and shared values into each context. A deterministic index-based rule equalizes counts without consulting image scores.

Figure~\ref{fig:exchange} shows that mask preference depends on context. At 50 and 70\% sparsity, every context places its three highest-energy masks above its median score across all 12 masks. At 90\%, only half do. The original four-model comparison gives a supported reversal on M4Raw at 70\%: the SNIP mask wins in the SNIP context, while the RigL mask wins in the RigL context (Table~\ref{tab:context}). Each mask's energy and values are identical across these contexts. The reversal therefore limits a universal mask-only ranking, while remaining compatible with useful average energy associations.

To reduce confounding by pruning-method identity, we evaluate the three constructed masks from Eq.~\ref{eq:built} in two randomly pruned contexts per level. On M4Raw, the fitted slopes through the three energy-quality points are positive at 60 to 80\% sparsity (6.5 to 10.9 dB per unit energy fraction) and negative at 90\% ($-6.8$ and $-8.4$). One 90\% context favors the middle-energy mask. Figure~\ref{fig:energy} makes this change in the energy-quality relationship visible at fixed count and overlap.

\begin{table}[t]
\centering
\caption{ViT controls on M4Raw (dB). Top: original-mask advantage over the 11 foreign masks, averaged over 12 contexts per level, before and after one training epoch. Bottom: the 90\% scale-control comparison. Original-mask lead uses the same contrast. High-energy effect is the reported high-energy-minus-count-matched-random mask contrast. Scale factors use 32 unlabeled slices.}
\label{tab:scale}
\setlength{\tabcolsep}{7pt}
\begin{tabular}{@{}lrrr@{}}\toprule
Original-mask advantage & 50\% & 70\% & 90\%\\\midrule
Before retraining & 1.73 & 2.45 & 4.64\\
After one epoch & .95 & 1.35 & 1.85\\
\bottomrule\end{tabular}
\par\vspace{5pt}
\setlength{\tabcolsep}{5pt}
\begin{tabular}{@{}lrr@{}}\toprule
90\% scale control & Original-mask & High-energy\\
 & lead & effect\\\midrule
None & 4.64 & $-4.67$\\
Per matrix & 3.28 & $+.08$\\
Per channel & 3.60 & $+.35$\\
\bottomrule\end{tabular}
\end{table}

At 90\%, high-energy exchanged masks produce intermediate MLP outputs roughly ten times their size under the context's original mask. Matching each matrix's output scale to that reference, estimated from 32 unlabeled slices, changes the reported high-energy-minus-random effect from $-4.67$ to $+.08$ dB. Per-channel matching gives $+.35$ dB. Original-mask advantages remain 3.28 and 3.60 dB, respectively (Table~\ref{tab:scale}). This control implicates scale mismatch in the extreme-sparsity penalty, while compatibility with the trained context remains relevant. It uses an available original-mask network as its calibration reference.

One epoch of fixed-mask retraining at learning rate $2\times10^{-5}$ provides a further check across all 432 combinations. In random-mask contexts, the mean score of the three highest-energy masks relative to the context's median changes from $-1.40$ to $+.04$ dB at 90\%. Foreign-mask scores improve, but remain below the correspondingly retrained original-mask networks at all three levels (Table~\ref{tab:scale}). Relative to the original-mask score before retraining, the 90\% deficit is still .67 dB. Retraining thus removes the random-context reversal while retaining the original-mask advantage.

\section{Conclusion}
Sparsity is a storage descriptor, but it is insufficient to characterize reconstruction quality. In U-Net, operation count explains part of the observed variation, while equal-operation controls reveal additional first-convolution sensitivity. In ViT, retained pretrained energy is informative within sparsity levels, but the same masks can reverse their ranking across trained networks. Scale calibration and retraining further show how the energy effect interacts with output scale at extreme sparsity. Architecture-specific descriptors therefore help interpret pruning, while the value of a mask can still depend on the surrounding trained network.

Evidence is limited to selected checkpoints, one training seed and simulated single-channel acquisition. The tested networks omit explicit data-consistency steps, unlike model-based MRI reconstruction networks such as MoDL and variational networks~\cite{modl,variational_mri}. ViT exchanges reset pretrained values, reducing the selected original networks' PSNR by 1.5 to 6.1 dB before exchange. The conclusions concern these controlled states, and require validation across training runs and native multicoil reconstruction systems.

\clearpage
\nobalance
\ifdefined\buildextended
\section{Additional details}
This working version includes the full main paper. The following details preserve additional evidence from both source projects. They are outside the four-page ICASSP technical paper.

\subsection{Protocol}
Let $z_c$ denote the complex image from receiver coil $c$. The preprocessing combines its real and imaginary parts separately as $r=(\sum_c(\Re z_c)^2)^{1/2}$ and $s=(\sum_c(\Im z_c)^2)^{1/2}$. With Fourier transform $\mathcal F$, sampling mask $P$ and imaginary unit $i$, the input is
\begin{equation}
 x=\mathcal N\!\left(\left|\mathcal F^{-1}\bigl(P\odot\mathcal F(r+is)\bigr)\right|\right),
\end{equation}
where $\mathcal N(v)=(v-\min v)/(\max v-\min v+10^{-8})$. The magnitude target is normalized independently. This makes the evaluated task a constructed single-channel reconstruction problem.

The noncentral sampling lines are selected by ranking uniform random scores using NumPy RandomState with seed 42. Jitter uses seed 1234 and can also move central lines. Rejected collisions and out-of-range positions preserve the number of lines. This operation changes sampling locations, without simulating physical motion. Each condition uses the first 1,500 eligible cached slices in file and slice order. Eligible indices are 5 through $n-3$ for fastMRI and 2 through $n-3$ for M4Raw, where $n$ is the number of slices. Unreadable files and slices are skipped. Training and test volume lists are disjoint.

The U-Net starts with 32 channels and uses dropout probability .1. ViT uses 16 attention heads per block, MLP width 2,816, and no dropout. Source training uses a single-cycle linear learning-rate schedule, updated each epoch. RigL updates connectivity every 2,335 batches until 75\% of training, starting with a replaced fraction of .1 for U-Net and .31 for ViT and decreasing it with a cosine schedule. Mask selection uses squared error on fully sampled normalized inputs for U-Net and on inputs from the ViT sampling pipeline for ViT. The five SNIP masks use five distinct source slices.

The recorded ViT PUN-IT search uses 500 steps, learning rate .05, Kullback-Leibler penalty weight 10, temperature decreasing from 5 to .1, and 200 examples in batches of two. Gate-noise standard deviation is .5. U-Net code defaults use batches of four, but its exact historical search configuration is unavailable. The common ViT reference has SHA256 prefix \texttt{00f1b7c250abe0f7}.

Bootstrap resampling keeps all slices of a sampled volume or study group together, with paired model scores. The 23 M4Raw study-ID groups contain 108 volumes and need not correspond uniquely to patients. Means preserve the original slice weighting. For a family of 16 comparisons, interval endpoints use bootstrap percentiles $100(.05/32)$ and $100(1-.05/32)$, targeting nominal 95\% simultaneous coverage. Three U-Net families and one ViT family are adjusted separately. These intervals condition on fixed checkpoints, masks and analysis choices. They do not adjust for selecting exploratory analyses after inspecting outputs.

\subsection{U-Net controls}
The equal-count rules delete $\lfloor .1N_{\rm body}\rfloor$ active weights, where $N_{\rm body}$ counts active entries in 18 body convolutions. An output channel with $N_{\rm row}$ active filter weights permits at most $\lfloor .2N_{\rm row}\rfloor$ deletions. Paired random priorities use NumPy default\_rng seeds 2026091400 through 2026091404. Under these constraints, H preserves the most theoretical operations and L the fewest.

Table~\ref{tab:extended_unet} gives the original 70\% M4Raw means. Protecting the first convolution moves 18, 0, 30 and 32 deletions for PUN-IT, random, RigL and SNIP, respectively. The operation-count increases relative to L are .046\%, 0\%, .044\% and .049\%. The corresponding $L_1-L$ gains are 1.967, 0, 1.555 and 2.325 dB. The nonzero gains have adjusted intervals [1.849, 2.081], [1.419, 1.691] and [2.154, 2.499] dB. This exploratory comparison forms its own 16-comparison family.
\begin{table}[t]
\centering
\caption{M4Raw PSNR (dB) in the initial U-Net experiment at 70\% sparsity, averaged over five deletion seeds. H preserves high-resolution computation, L deletes it first, and $L_1$ protects the first convolution within L.}
\label{tab:extended_unet}
\setlength{\tabcolsep}{6pt}
\begin{tabular}{@{}lrrrr@{}}\toprule
Model & Original & H & L & $L_1$\\\midrule
PUN-IT & 27.62 & 27.63 & 25.00 & 26.97\\
Random & 27.71 & 27.70 & 26.85 & 26.85\\
RigL & 27.92 & 27.86 & 25.99 & 27.54\\
SNIP & 28.04 & 28.17 & 25.43 & 27.76\\
\bottomrule\end{tabular}
\end{table}

The squared weight magnitudes removed by L are 1.04 to 1.21 times those removed by H in the initial study. Counting valid spatial uses raises this ratio to 11.88 to 16.96. For a convolution represented by matrix $A_l$, deletion set $D_l$ and valid-use count $t_{lj}$ for weight $w_{lj}$, this follows from the exact identity
\begin{equation}
 \sum_l\|\Delta A_l\|_F^2=\sum_{l,j\in D_l}t_{lj}w_{lj}^2.
\end{equation}
Boundary-padding zeros do not contribute valid uses. This identity describes the edited linear operator. A local feature response also depends on the input features, which motivates measuring that response directly.

For fixed reference input patches $q_{ilp}$ at layer $l$, image $i$ and spatial position $p$, let $K_l$ contain removed filter weights and define the uncentered second-moment matrix $C_l=n^{-1}\sum_{i,p}q_{ilp}q_{ilp}^{\top}$. The image-averaged squared local output change is
\begin{equation}
 Q_l=\frac1n\sum_{i,p}\|K_lq_{ilp}\|^2
 =\operatorname{tr}(K_lC_lK_l^{\top}).
\end{equation}
Positions are summed and images averaged. A diagonal $C_l$ weights input features by their separate squared strengths. The stronger condition $C_l=c_lI$ yields $Q_l=c_l\|K_l\|_F^2$. Thus deleted weight energy predicts this local response only under additional conditions on input features.

The recorded measurements use outputs after instance normalization, which centers and rescales feature channels. Their L-to-H squared-output-change ratios are 3.89 to 7.82, or 1.70 to 5.62 after excluding the first layer from the summary. These ratios exceed one in all 80 paired method-condition-seed comparisons. Excluding a layer changes the diagnostic summary, while the PSNR comparison still evaluates each complete edited network. H exceeds $L_1$ in 15 of 16 adjusted comparisons. The unresolved case is SNIP on fastMRI with jitter, with difference .023 dB and interval [$-.032,.078$].

\subsection{ViT controls}
The initial four-model exchange equalizes masks at 11,894,784 active MLP weights. It removes 0 random-mask entries, 1 SNIP entry, 3 PUN-IT entries and 16 RigL entries. The deterministic procedure traverses parameter arrays in reverse alphabetical order and removes active entries with the largest flattened indices. Table~\ref{tab:extended_exchange} retains the full M4Raw mean matrix. The mask energy fractions are 63.39\%, 29.99\%, 41.35\% and 40.31\% for SNIP, random, PUN-IT and RigL. Every entry uses 94.935 GFLOPs.
\begin{table}[t]
\centering
\caption{ViT M4Raw PSNR at 70\% sparsity with shared pretrained MLP values. Rows choose the trained context. Columns choose the mask. Diagonal entries also reset MLP values.}
\label{tab:extended_exchange}
\setlength{\tabcolsep}{5pt}
\begin{tabular}{@{}lrrrr@{}}\toprule
Context & SNIP & Random & PUN-IT & RigL\\\midrule
SNIP & 27.085 & 19.260 & 18.693 & 22.842\\
Random & 23.520 & 20.976 & 18.020 & 23.696\\
PUN-IT & 23.279 & 19.316 & 22.526 & 22.766\\
RigL & 22.950 & 21.900 & 17.692 & 23.424\\
\bottomrule\end{tabular}
\end{table}

The original trained PSNR values in the same row order are 28.557, 27.117, 26.374 and 27.019 dB. Resetting the retained MLP values before count equalization lowers them by 1.473, 6.141, 3.848 and 3.596 dB. Reset scores differ from the equalized diagonal values by less than .001 dB. This reset cost scopes the exchange result to the constructed networks.

The SNIP-minus-RigL ordering also reverses on fastMRI. In the SNIP context, the difference is 7.444 dB with adjusted interval [7.036, 7.833]. In the RigL context it is $-.771$ dB with interval [$-1.054,-.486$], and in the random context it is $-.588$ dB with interval [$-.882,-.296$]. The negative RigL-context differences under jitter are unresolved: $-.139$ dB on fastMRI and $-.107$ dB on M4Raw, with intervals [$-.366,.085$] and [$-.409,.246$]. This distinction separates supported reversals from changes in point-estimate signs.

Table~\ref{tab:extended_energy} reports the trained-model energy correlations under all four conditions. Those associations compare trained models, whereas the exchanges fix a common set of pretrained MLP values. Consequently, the exchange demonstrates a limit on universal energy-based ranking without establishing a complete explanation for the trained-model correlations.
\begin{table}[t]
\centering
\caption{Pearson correlation between retained pretrained energy and PSNR across 60 trained ViTs. Within-level ranges cover five levels, each with 12 models. Jitter changes sampling-line positions.}
\label{tab:extended_energy}
\setlength{\tabcolsep}{7pt}
\begin{tabular}{@{}lcc@{}}\toprule
Condition & Pooled & Within level\\\midrule
fastMRI & .88 & .81 to .97\\
fastMRI + jitter & .88 & .73 to .97\\
M4Raw & .83 & .77 to .96\\
M4Raw + jitter & .80 & .68 to .91\\
\bottomrule\end{tabular}
\end{table}

For the three constructed masks with fixed count and zero overlap, the fitted slopes in the two random contexts are 10.9 and 6.5 at 60\%, 10.5 and 8.2 at 70\%, 8.1 and 7.1 at 80\%, and $-6.8$ and $-8.4$ at 90\%. Units are dB per unit retained-energy fraction. These fits use only three points each. Zero overlap constrains similarity to the original support, while the construction still changes selected positions and potentially layer counts.

Table~\ref{tab:extended_retrain} separates two retraining effects. The high-energy preference is the mean PSNR of a context's three highest-energy masks minus the median over its 12 masks, averaged over the stated context family. Foreign-mask recovery is the mean over 132 off-diagonal combinations per sparsity level relative to the original-mask score before retraining. The main manuscript separately reports deficits relative to an original-mask network retrained for the same epoch.
\begin{table}[t]
\centering
\caption{ViT M4Raw summaries before and after one fixed-mask training epoch. Entries are in dB. High-energy preference uses five contexts per family. Foreign-mask recovery uses 132 combinations per level and the original-mask score before retraining as reference.}
\label{tab:extended_retrain}
\setlength{\tabcolsep}{5pt}
\begin{tabular}{@{}llrrr@{}}\toprule
Summary & Epoch & 50\% & 70\% & 90\%\\\midrule
SNIP & Before & 3.53 & 5.38 & 8.25\\
preference & After & 1.78 & 2.69 & 4.74\\
Random & Before & .71 & .44 & $-1.40$\\
preference & After & .23 & $-.05$ & .04\\
Foreign & Before & $-1.73$ & $-2.45$ & $-4.64$\\
recovery & After & 2.48 & 1.33 & $-.67$\\
\bottomrule\end{tabular}
\end{table}

Additional restoration checks add pretrained values at inactive MLP positions while keeping learned active values and context fixed. Full-strength additions reduce PSNR by 4.360 to 9.023 dB across 16 original model-condition comparisons. Quarter-strength additions reduce it in 15 comparisons, with a .072 dB gain for SNIP on fastMRI with jitter. These are descriptive checks that change active count and are evaluated before retraining.

\clearpage
\fi

\section{Compliance with Ethical Standards}
This retrospective analysis used released anonymized fastMRI and M4Raw data, with no new participants or acquisitions. fastMRI curation was approved by the New York University School of Medicine Institutional Review Board~\cite{fastmri}. M4Raw acquisition was approved by Shenzhen Technology University's Institutional Review Board (SZTULL012021005), with written consent for anonymized public release~\cite{m4raw}.

\section{Acknowledgments}
No funding was received for conducting this study. The authors have no relevant financial or nonfinancial interests to disclose.

\bibliographystyle{IEEEbib}
\bibliography{references,additional_references}

@article{ssim,
  author={Zhou Wang and Alan C. Bovik and Hamid R. Sheikh and Eero P. Simoncelli},
  title={Image Quality Assessment: From Error Visibility to Structural Similarity},
  journal={IEEE Transactions on Image Processing},
  volume={13}, number={4}, pages={600-612}, year={2004},
  doi={10.1109/TIP.2003.819861}
}

@inproceedings{blalock,
  title={What is the state of neural network pruning?},
  author={Blalock, Davis and Gonzalez Ortiz, Jose Javier and Frankle, Jonathan and Guttag, John},
  booktitle={Proceedings of Machine Learning and Systems (MLSys)},
  year={2020}
}

@inproceedings{unet,
  author    = {Olaf Ronneberger and Philipp Fischer and Thomas Brox},
  title     = {{U-Net}: Convolutional Networks for Biomedical Image Segmentation},
  booktitle = {MICCAI},
  year      = {2015}
}

@article{fastmri,
  author  = {Jure Zbontar and others},
  title   = {{fastMRI}: An Open Dataset and Benchmarks for Accelerated {MRI}},
  journal = {arXiv preprint arXiv:1811.08839},
  year    = {2018}
}

@article{m4raw,
  author  = {Mengye Lyu and others},
  title   = {{M4Raw}: A Multi-Contrast, Multi-Repetition, Multi-Channel {MRI} K-Space Dataset for Low-Field {MRI} Research},
  journal = {Scientific Data},
  volume  = {10},
  pages   = {264},
  year    = {2023},
  doi     = {10.1038/s41597-023-02181-4}
}

@inproceedings{snip,
  author    = {Namhoon Lee and Thalaiyasingam Ajanthan and Philip H. S. Torr},
  title     = {{SNIP}: Single-Shot Network Pruning Based on Connection Sensitivity},
  booktitle = {ICLR},
  year      = {2019}
}

@inproceedings{rigl,
  author    = {Utku Evci and Trevor Gale and Jacob Menick and Pablo Samuel Castro and Erich Elsen},
  title     = {Rigging the Lottery: Making All Tickets Winners},
  booktitle = {ICML},
  year      = {2020}
}

@inproceedings{punit,
  author    = {Shijun Liang and Evan Bell and Avrajit Ghosh and Saiprasad Ravishankar},
  title     = {Pruning Unrolled Networks ({PUN}) at Initialization for {MRI} Reconstruction Improves Generalization},
  booktitle = {Asilomar Conference on Signals, Systems, and Computers},
  pages     = {100-104},
  year      = {2024},
  doi       = {10.1109/IEEECONF60004.2024.10942630}
}

@inproceedings{vit,
 author={Kang Lin and Reinhard Heckel},
 title={Vision Transformers Enable Fast and Robust Accelerated {MRI}},
 booktitle={Medical Imaging with Deep Learning},
 pages={774-795}, year={2022}
}

@inproceedings{lottery,
 author={Jonathan Frankle and Michael Carbin},
 title={The Lottery Ticket Hypothesis: Finding Sparse, Trainable Neural Networks},
 booktitle={ICLR}, year={2019}}

@inproceedings{missing,
 author={Jonathan Frankle and Gintare Karolina Dziugaite and Daniel M. Roy and Michael Carbin},
 title={Pruning Neural Networks at Initialization: Why Are We Missing the Mark?},
 booktitle={ICLR}, year={2021}}

@inproceedings{grasp,
 author={Chaoqi Wang and Guodong Zhang and Roger Grosse},
 title={Picking Winning Tickets Before Training by Preserving Gradient Flow},
 booktitle={ICLR}, year={2020}}

@inproceedings{synflow,
 author={Hidenori Tanaka and Daniel Kunin and Daniel L. K. Yamins and Surya Ganguli},
 title={Pruning Neural Networks Without Any Data by Iteratively Conserving Synaptic Flow},
 booktitle={NeurIPS}, year={2020}}

@inproceedings{robustness,
 author={Mohammad Zalbagi Darestani and Akshay S. Chaudhari and Reinhard Heckel},
 title={Measuring Robustness in Deep Learning Based Compressive Sensing},
 booktitle={ICML}, pages={2433-2444}, year={2021}}

@article{antun,
 author={Vegard Antun and Francesco Renna and Clarice Poon and Ben Adcock and Anders C. Hansen},
 title={On Instabilities of Deep Learning in Image Reconstruction and the Potential Costs of {AI}},
 journal={Proceedings of the National Academy of Sciences},
 volume={117}, number={48}, pages={30088-30095}, year={2020}, doi={10.1073/pnas.1907377117}}

@article{modl,
 author={Hemant K. Aggarwal and Merry P. Mani and Mathews Jacob},
 title={{MoDL}: Model-Based Deep Learning Architecture for Inverse Problems},
 journal={IEEE Transactions on Medical Imaging}, volume={38}, number={2},
 pages={394-405}, year={2019}, doi={10.1109/TMI.2018.2865356}}

@inproceedings{rethinking,
 author={Zhuang Liu and Mingjie Sun and Tinghui Zhou and Gao Huang and Trevor Darrell},
 title={Rethinking the Value of Network Pruning},
 booktitle={ICLR}, year={2019}}

@article{variational_mri,
 author={Kerstin Hammernik and Teresa Klatzer and Erich Kobler and Michael P. Recht and Daniel K. Sodickson and Thomas Pock and Florian Knoll},
 title={Learning a Variational Network for Reconstruction of Accelerated {MRI} Data},
 journal={Magnetic Resonance in Medicine}, volume={79}, number={6},
 pages={3055-3071}, year={2018}, doi={10.1002/mrm.26977}}

@article{mri_generalization,
 author={Florian Knoll and Kerstin Hammernik and Erich Kobler and Thomas Pock and Michael P. Recht and Daniel K. Sodickson},
 title={Assessment of the Generalization of Learned Image Reconstruction and the Potential for Transfer Learning},
 journal={Magnetic Resonance in Medicine}, volume={81}, number={1},
 pages={116-128}, year={2019}, doi={10.1002/mrm.27355}}

@article{wattad_kspace,
 author={Mohammed Wattad and Tamir Shor and Alex Bronstein},
 title={On the Role of {K-Space} Acquisition in {MRI} Reconstruction Domain-Generalization},
 journal={arXiv preprint arXiv:2512.06530}, year={2025},
 doi={10.48550/arXiv.2512.06530}}

@inproceedings{lamp,
 author={Jaeho Lee and Sejun Park and Sangwoo Mo and Sungsoo Ahn and Jinwoo Shin},
 title={Layer-Adaptive Sparsity for the Magnitude-Based Pruning},
 booktitle={ICLR}, year={2021}}

@inproceedings{wanda,
 author={Mingjie Sun and Zhuang Liu and Anna Bair and J. Zico Kolter},
 title={A Simple and Effective Pruning Approach for Large Language Models},
 booktitle={ICLR}, year={2024}}

@inproceedings{obd,
 author={Yann LeCun and John S. Denker and Sara A. Solla},
 title={Optimal Brain Damage},
 booktitle={Advances in Neural Information Processing Systems},
 volume={2}, pages={598-605}, year={1989}}
\end{document}